\documentclass[a4paper,12pt]{article}
\usepackage[latin1]{inputenc}
\usepackage{float}
\usepackage{amsmath,amssymb}
\usepackage{theorem}
\usepackage{graphicx}

\begin{document}

\title{\bf Perturbations with angular momentum of Robinson-Trautman spacetimes}
\author{Osvaldo M. Moreschi${}^1$\thanks{Member of CONICET.}
\thanks{ Email: moreschi@fis.uncor.edu}~
and
Alejandro P\'{e}rez${}^2$\thanks{
present address:  
Centre de Physique Th\'eorique, CNRS Luminy,  
             F-13288 Marseille, France. 
}
\thanks{Andrew Mellon Predoctoral Fellowship holder.} \\
${}^1$\small FaMAF, Universidad Nacional de C\'{o}rdoba\\
\small Ciudad Universitaria,
(5000) C\'{o}rdoba, Argentina.\\
${}^2$\small Department of Physics and Astronomy,
University of Pittsburgh,  \\
\small 100 Allen Hall,
Pittsburgh, PA 15260, USA.
}
\date{March 8, 2001}
\maketitle

\begin{abstract}
We study the possible asymptotically flat perturbations of Robinson-Trautman
spacetimes. We differentiate between algebraically special perturbations 
and general perturbations.
The equations that determine physically realistic spacetimes with
angular momentum are presented.
\end{abstract}


\section{Introduction}

Robinson-Trautman\cite{Robinson62} spacetimes (RT) have been very useful 
for estimating the total gravitational radiation in the head-on black hole
collision\cite{Moreschi96}\cite{Moreschi99}\cite{Anninos93}. 
In reference \cite{Moreschi96} we have applied these geometries to the 
description of
the total energy radiated in the head-on black hole collision with equal
mass; and it was shown that our calculations agree remarkably well with
the numerical exact calculations of Anninos et.al.\cite{Anninos93}.
The case of unequal mass black hole collision, was treated numerically
in reference \cite{Anninos98}; and our technique based on the use
of the RT geometries\cite{Moreschi99} showed again an impressive 
agreement with the exact calculations.
If one wants to generalize these estimates to the case of
the black hole collision with orbital angular momentum it is necessary to
consider spacetimes with total angular momentum. This is the main
physical motivation for this work; the application of these calculations
to specific physical models will be presented elsewhere\cite{Moreschi01}.

The Robinson-Trautman vacuum solutions are algebraically special spacetimes
which are characterized by the existence of 
a congruence of diverging null geodesics without shear and twist. This
implies the existence of a prefered family of null hypersurfaces; 
which provides a set of sections at future null infinity. 
The angular momentum calculated on these sections is found to be
zero. (In a Bondi frame the asymptotic NP quantities $\Psi_1^0$ and
$\sigma_0$ are zero; as a consequence the angular momentum vanishes
independently of the definition used. See for example reference 
\cite{Moreschi86} and references therein.) In section \ref{sec:RT}
we present a short review of the RT vacuum solutions.

It is natural then to consider the algebraically special perturbations of RT
spacetimes; since they contain the RT family in the limit when the twist
goes to zero. 

Algebraically special spacetimes are physically very important solutions
since they 
contain the RT family, the Kerr family of
solutions and so also the Schwarzschild solution.
In section \ref{sec:algsp} we present in full detail the algebraically 
special vacuum solutions in terms of a new frame,
the so called `standard rotating null tetrad'\cite{Perez01}.

The algebraically special perturbations of RT spacetimes are
governed by rather complicated equations;
however, the equations simplify when one analysis the geometry in the 
far future in the limit of the asymptotic retarded time going to 
infinity. This is due to the fact that the RT geometry 
simplifies considerably in this limit.
It will be shown that the total angular momentum of these perturbations
is constant in the asymptotic future; therefore these spacetimes are not
suitable for the description of a perturbed black hole with a
dynamic angular momentum. 
Furthermore, the general algebraically special perturbations of RT spacetimes
turn out to be divergent in the asymptotic future.
These characteristics do not seem to be appropriate for the modeling
of a spacetime representing
the last stage after the collision 
of two compact objects with non-zero total angular momentum.
We are therefore forced to
consider general perturbations of RT spacetimes.

We dedicate section \ref{sec:general} to the presentation of
general perturbations of RT spacetimes. The radial dependence of the
tetrad is explicitly solved; and we also present the intrinsic
evolution equations at future null infinity, which are 
used in the calculation of gravitational radiation.

The important issue of gauge freedom is discussed in both cases,
algebraically special spacetimes (section \ref{sec:algsp}) and 
general perturbation of RT spacetimes (section \ref{sec:general}).

Summarizing comments are included in the last
section \ref{sec:final}.

Throughout the work we will use the GHP\cite{Geroch73} notation.

\section{Robinson-Trautman spacetimes}\label{sec:RT}

Robinson and Trautman[1] studied the vacuum solutions containing a
congruence of diverging null geodesics, with vanishing shear and twist. The
line element of these metrics can be expressed[4] by:
\begin{equation}
ds^{2}=\left( -2Hr+K-2\frac{M(u)}{r}\right) du^{2}+2\;du\;dr-\frac{r^{2}}{%
P^{2}}d\zeta \;d\bar{\zeta} , \label{eqRTlinelement}
\end{equation}
where $P=P(u,\zeta ,\bar{\zeta})$, $H=\frac{\dot{P}}{P}$, $K=\Delta \ln P$
, a doted quantity denotes its time derivative,
a bar means complex conjugate
 and $\Delta$ is the
two-dimensional Laplacian for the two-surfaces $u=\tt constant$, $r=\tt constant$ 
with line element 
\begin{equation}\label{eq:desphere}
dS^{2}=\frac{1}{P^{2}}\;d\zeta \;d\bar{\zeta} ;
\end{equation}
where we are using complex stereographic coordinates $(\zeta, \bar\zeta)$
for the sphere.

It is usually convenient to describe this line element in terms of the line
element of the unit sphere; this is done by expressing $P$ as the product 
$P=V(u,\zeta ,\bar{\zeta})P_{0}(\zeta ,\bar{\zeta})$, where $P_{0}$ is the
value of $P$ for the unit sphere. If $\ell$ denotes the vector field that generates
the null congruence, then $\ell =du$, $\ell (r)=1$, $\ell (\zeta )=0$ and $%
\ell (\bar{\zeta})=0$. In other words this is the coordinate system adapted
to the geometry. It is convenient to use the parameterization $u$ such that 
the mass parameter $M(u)=M_{0}=\tt constant$.

Alternatively one can express the geometry in terms of a null tetrad; since,
given the complex null vector basis $\left( \ell ^{a},m^{a},\bar{m}%
^{a},n^{a}\right) $ with the properties:
\begin{equation}\label{eq:produc}
g_{ab}\;\ell ^{a}\;n^{b}=-g_{ab}\;m^{a}\;\bar{m}^{b}=1
\end{equation}
and all other possible scalar products being zero, the metric can be
expressed by
\begin{equation}
g_{ab}=\ell _{a}\;n_{b}+n_{a}\;\ell _{b}-m_{a}\;\bar{m}_{b}-\bar{m}%
_{a}\;m_{b}.
\end{equation}
Using the coordinate system $(x^{0},x^{1},x^{2},x^{3})=\left( u,r,%
(\zeta +\bar{\zeta}),\frac{1}{i}(\zeta -\bar{\zeta})\right)$
defined above, one can express the null tetrad as:
\begin{equation}
\ell _{a}=\left( du\right)_{a} 
\label{uno}
\end{equation}
\begin{equation}
\ell ^{a}=\left( \frac{\partial }{\partial \,r}\right) ^{a} 
\label{dos}
\end{equation}
\begin{equation}
m^{a}=\xi ^{i}\left( \frac{\partial }{\partial x^{i}}\right) ^{a} 
\end{equation}
\begin{equation}
\bar{m}^{a}=\bar{\xi}^{i}\left( \frac{\partial }{\partial x^{i}}\right) ^{a} 
\label{tres}
\end{equation}
\begin{equation}\label{eq:vecn}
n^{a}=
 \,U\,\left( \frac{\partial }{\partial \,r}\right)^{a}
+ X^{i}\,\left(\frac{\partial }{\partial \,x^{i}}\right)^{a} 
\end{equation}
with $i=0,2,3$  and
where the components $\xi^{i}$, $U$ and $X^{i}$ are:
\begin{equation}
\xi ^{0}=0,\quad \xi ^{2}=\frac{\xi _{0}^{2}}{r},\quad \xi ^{3}=\frac{\xi
_{0}^{3}}{r}, 
\end{equation}
with
\begin{equation}\label{eq:xileading}
\xi _{0}^{2}=\sqrt{2}P_{0}\;V,\qquad \xi _{0}^{3}=-i\xi _{0}^{2}; 
\end{equation}
\begin{equation}
U=rU_{00}+U_{0}+\frac{U_{1}}{r}, 
\end{equation}
where
\begin{equation}
U_{00}=\frac{\dot{V}}{V},\quad U_{0}=-\frac{1}{2}K_{V},\quad U_{1}=-\frac{%
\Psi _{2}^{0}+\bar{\Psi}_{2}^{0}}{2}, 
\end{equation}
where the curvature $K_V$ of the 2-metric appearing in equation 
(\ref{eq:desphere}), is given by
\begin{equation}
K_{V}=\frac{2}{V}~\bar{\eth }_{V}\eth _{V}\, V-\frac{2}{V^{2}}~\eth _{V}V~%
\bar{\eth }_{V}V+V^{2}, 
\end{equation}
the leading order behavior $\Psi_2^0$ of the second component of the Weyl
tensor is $\Psi_{2}^{0}=-M_{0}$
and
\begin{equation}
X^{0}=1,\quad X^{2}=0,\quad X^{3}=0;
\end{equation}
and where $\eth_V$ is the edth operator, in the GHP notation, 
of the sphere with metric 
(\ref{eq:desphere}).

In reference [1] it was found that the vacuum Einstein equation can be
reduced to a parabolic equation for a scalar depending on three variables,
the so called Robinson-Trautman equation; which in our notation has the form
\begin{equation}
-3\, M_{0}\, \dot{V}=V^{4}~\eth ^{2}\bar{\eth }^{2}~V-V^{3}~\eth ^{2}V~%
\bar{\eth }^{2}V;  \label{eqRTequation}
\end{equation}
where $\eth $ is the edth operator of the unit sphere. We refer to a line
element with $V$ satisfying this equation as a Robinson-Trautman solution. On
the other hand, if the RT equation is not required, then the solution is no
longer vacuum and there is only one component of the Ricci tensor different
from zero, given by\cite{Dain96}
\begin{equation}
\Phi _{22}^{(RT)}=\frac{-3\, M_{0}\frac{\dot{V}}{V}~-V^{3}~\eth ^{2}\bar{%
\eth }^{2}~V+V^{2}~\eth ^{2}V~\bar{\eth }^{2}V}{r^{2}};  \label{eqRTRicci}
\end{equation}
where the {\it (RT)} is to emphasize the fact that in this case we are using
the null tetrad adapted to the null congruence. We refer to this as a
Robinson-Trautman geometry.

At this point it is interesting to remark that the spin coefficients 
$\sigma, \tau, \kappa, \sigma', \tau'$ and $\epsilon$ are zero in the
Robinson-Trautman geometries.

\section{Twisting algebraically special spacetimes}\label{sec:algsp}

\subsection{Null tetrad components of the exact solutions}\label{sec:asexact}

As it was indicated previously, from a geometrical point of view, the
twisting algebraically special spacetimes are the natural generalizations of
RT spacetimes, since they also have a congruence of null geodesics with
vanishing shear and contain the RT metrics as the limit of those spaces when
the twist goes to zero. 

This kind of solutions have been extensively studied in the literature
in terms of different tetrads and coordinate systems (for an account of 
these formulations of the problem see \cite{Kramer80} and references therein). 
The tetrad and coordinate system used here
do not agree with those ones. This new tetrad corresponds to what 
we call ``standard rotating null tetrad'';
they are defined in reference \cite{Perez01}, and they can be applied to
any asymptotically flat spacetime. In this section we will 
express the equations governing twisting algebraically special spacetimes
in terms of the ``standard rotating null tetrad''.

The null tetrad components defined in equations (\ref{dos}) to (\ref{eq:vecn})  
can be expressed by the three 
scalars $M(u,\zeta,\bar\zeta)$, $V(u,\zeta,\bar\zeta)$ 
and $L(u,\zeta,\bar\zeta)$,
and are given by:
\begin{equation}
\xi ^{0}=\frac{\xi _{0}^{0}}{r-ic},\quad \xi ^{2}=\frac{\xi _{0}^{2}}{r-ic}%
,\quad \xi ^{3}=\frac{\xi _{0}^{3}}{r-ic}, 
\end{equation}
where
\begin{equation}
\xi _{0}^{0}=L\;V,\quad \xi _{0}^{2}=\sqrt{2}P_{0}\;V,\quad \xi
_{0}^{3}=-i\xi _{0}^{2},\quad P_{0}=\frac{1+\zeta \bar{\zeta}}{2};
\end{equation}
\begin{equation}
U=r\;U_{00}+U_{0}+\frac{\left( M\;r+\mu\;c\right) }{\left( r+ic\right)
\left( r-ic\right) }-\left( \frac{\tau _{0}\bar{\tau}_{00}}{r-ic}+\frac{\bar{%
\tau}_{0}\tau _{00}}{r+ic}\right) -\frac{\tau _{0}\;\bar{\tau}_{0}}{\left(
r+ic\right) \left( r-ic\right) }, 
\end{equation}
where
\begin{equation}
U_{00}=\frac{\dot{V}}{V},\quad U_{0}=-\frac{1}{2}\left( K_{V}+K_{L}\right) , 
\end{equation}
\begin{equation}
K_{V}=2V~\bar{\eth }\eth ~V-2~\eth V~\bar{\eth }V+V^{2}, 
\end{equation}
\begin{multline}
K_{L} =2\dot{\bar{L}} \dot{L}V^2
+\left( \dot L \bar L V + \dot{\bar{L}}L V-2\eth V\bar{%
L}+V\eth \bar{L}+V\bar{\eth }L-2\bar{\eth }V~L\right) \dot{V} \\
+V^{2}\ddot{L}\bar{L}+2\ddot{V}V L\bar{L}-2%
\dot{V}^{2}L\bar{L}+V^{2}L\ddot{\bar{L}} \\
+2\eth \dot{V}~V(L+\bar{L})+V^{2}(\eth \bar{L}+\bar{%
\eth }L)
\end{multline}
\begin{equation}
\begin{split}
\mu =
& \ddot c\,L\,V^2\,\bar L + 2\,\dot c\,\dot L\,V^2\,\bar L 
+ \dot c\,\dot{\bar L}  \,L\,V^2 + \dot c\,\bar\eth L \,V^2 \\
&+ \dot c\,c\,i + \frac{1}{2}\,\ddot L\,c\,V^2\,\bar L
+ \dot L\,\bar\eth c \,V^2 +  \frac{1}{2}\,\dot L\,\eth c \,L^{-1}\,V^2\,\bar L 
+ \dot L\,c^2\,i\,L^{-1} \\
& +  \frac{1}{2}\,\ddot{\bar L}\,c\,L\,V^2 +  \frac{1}{2}\,\dot{\bar L} \,\eth c\,V^2 
+ \bar\eth(\dot c)\,L\,V^2
 +  \frac{1}{2}\,\bar\eth \dot L \,c\,V^2 + \bar\eth \eth c\,V^2 \\
&+  \frac{1}{2}\,\bar\eth L \,\eth c \,L^{-1}\,V^2 + \eth \dot c \,V^2\,\bar L 
+  \frac{1}{2}\,\eth \dot{\bar L}  \,c\,V^2 \\
&-  \frac{1}{2}\,\eth c \,\eth \bar L \,L^{-1}\,V^2 + c\,\eth c \,i\,L^{-1} 
+ c(K_V + K_L)
\end{split}
\end{equation}
\begin{equation}
\tau _{00}=-\dot{L}V, 
\end{equation}
\begin{equation}
\tau _{0}=2ic\tau _{00}-iV\eth _{L}c, 
\end{equation}
\begin{equation}
c=-\frac{V^{2}}{2i}\left( \bar{\eth }_{L}L-\eth _{L}\bar{L}\right) ;
\end{equation}
\begin{equation}
X^{0}=1-\frac{\bar{\tau}_{00}\xi _{0}^{0}}{r-ic}-\frac{\tau _{00}\bar{\xi}%
_{0}^{0}}{r+ic}-\frac{\bar{\tau}_{0}\xi _{0}^{0}+\tau _{0}\bar{\xi}_{0}^{0}}{%
\left( r+ic\right) \left( r-ic\right) }, 
\end{equation}
\begin{equation}
X^{2}=-\frac{\bar{\tau}_{00}\xi _{0}^{2}}{r-ic}-\frac{\tau _{00}\bar{\xi}%
_{0}^{2}}{r+ic}-\frac{\bar{\tau}_{0}\xi _{0}^{2}+\tau _{0}\bar{\xi}_{0}^{2}}{%
\left( r+ic\right) \left( r-ic\right) }, 
\end{equation}
\begin{equation}
X^{3}=-\frac{\bar{\tau}_{00}\xi _{0}^{3}}{r-ic}-\frac{\tau _{00}\bar{\xi}%
_{0}^{3}}{r+ic}-\frac{\bar{\tau}_{0}\xi _{0}^{3}+\tau _{0}\bar{\xi}_{0}^{3}}{%
\left( r+ic\right) \left( r-ic\right) };
\end{equation}
where the operator $\eth_{L}$ acting on any function $f=f(u,\zeta ,\bar{\zeta%
})$ is defined by:
\begin{equation}
\eth _{L}f\equiv \eth f+L\;\dot{f} 
\end{equation}
and
\begin{equation}
\bar{\eth}_{L}f\equiv \bar{\eth }f+\bar{L}\;\dot{f}. 
\end{equation}

The spin coefficients corresponding to this tetrad are:
\begin{equation}
  \label{eq:rho}
  \rho=\frac{-1}{r+ic},
\end{equation}
\begin{equation}
  \label{eq:sig}
  \sigma=0,
\end{equation}
\begin{equation}
  \label{eq:kaa}
  \kappa=0,
\end{equation}
\begin{equation}
  \label{eq:tau}
  \tau=\frac{\tau_{00}}{r+ic}+\frac{\tau_0}{(r+ic)(r-ic)},
\end{equation}

\begin{equation}
  \label{eq:rhop}
  \rho'=\frac{\rho'_0}{r-ic}-
\frac{V\, \eth_L \tau'_0-\eth_L V \, \tau'_0}{(r+ic)(r-ic)}+
\frac{\Psi_2^0 \, r}{(r-ic)^2(r+ic)} +
\frac{iV\, \tau'_0 \, \eth_L c}{(r+ic)^2(r-ic)} ,
\end{equation}
\begin{equation}
  \label{eq:sigp}
  \sigma'=\frac{\sigma'_0}{r+ic}-
\frac{V\, \bar\eth_L \tau'_0 + \bar\eth_L V \, \tau'_0}{(r+ic)^2}+
\frac{iV\, \tau'_0 \, \bar\eth_L c + (\tau'_0)^2}{(r+ic)^3},
\end{equation}
\begin{equation}
  \label{eq:kaap}
  \kappa'=\frac{V\, \bar\eth_L U}{r+ic} - U (\tau' +\bar\tau),
\end{equation}
\begin{equation}
  \label{eq:taup}
  \tau'= - \frac{\bar\tau_0}{(r+ic)^2},
\end{equation}
\begin{equation}
  \label{eq:eps}
  \epsilon=0,
\end{equation}
\begin{equation}
  \label{eq:epsp}
  \epsilon'=\epsilon'_0 + 
\frac{\beta_0 \bar\tau_{00}}{r-ic} -
\frac{\bar\beta_0 \tau_{00}}{r+ic} +
\frac{\beta_0 \bar\tau_0 - \bar\beta_0 \tau_0}{(r+ic)(r-ic)} +
\frac{\Psi_2^0 +2\bar\tau_0 \, \tau_{00}}{2(r+ic)^2} +
\frac{\tau_0 \bar\tau_0}{(r+ic)^2(r-ic)},
\end{equation}
\begin{equation}
  \label{eq:bet}
  \beta=\frac{\beta_0}{r-ic}
\end{equation}
\begin{equation}
  \label{eq:betp}
  \beta'=\bar\beta + \tau';
\end{equation}
where 
\begin{equation}\label{eq:rhop0}
\rho'_0= \frac{K_V+K_L}{2} - \dot L \dot{\bar{L}}\,V^2 +
\frac{\dot V}{V}ic -\frac{1}{2}(\bar\eth_L \dot L 
+\eth_L \dot{\bar{L}}) \, V^2,
\end{equation}
\begin{equation}
  \label{eq:sigp0}
  \sigma'_0=-\dot{\bar{L}}^2\,V^2 - \bar\eth_L(\dot{\bar{L}} V^2),
\end{equation}
\begin{equation}
  \label{eq:taup0}
  \tau'_0=-\bar\tau_0 ,
\end{equation}
\begin{equation}
  \label{eq:bet0}
  \beta_0=-\frac{\sqrt{2}V\left(\frac{\partial P_0}{\partial x^2}
-i\frac{\partial P_0}{\partial x^3} \right)+ \eth_L V }{2},
\end{equation}
\begin{equation}
  \label{eq:epsp0}
  \epsilon'_0=\frac{\dot V}{2V}.
\end{equation}

The Weyl tensor has the properties:
\begin{equation}
\Psi _{0}=0, 
\end{equation}
\begin{equation}
\Psi _{1}=0, 
\end{equation}
\begin{equation}
\Psi _{2}=\frac{\Psi_2^0}{(r+ic)^3}, 
\end{equation}
where
\begin{equation}
\Psi _{2}^{0}=-(M+i\mu).
\end{equation}
The components $\Psi_3$ and $\Psi_4$ are too complicated to be presented here.

The vacuum equations require several relations among the main functions. The 
equation $\Phi_{12}=0$ imposes the condition:
\begin{equation}\label{eq:phi120}
\eth_L (M+i\mu)=- 3\dot{L}(M+i\mu) .
\end{equation}
The other relation has a compact form if one uses the function 
$z(u,\zeta,\bar\zeta)$ defined by
\begin{equation}
  \label{eq:zeta}
  V=\frac{\partial z}{\partial u}=\dot z;
\end{equation}
using this, the equation $\Phi_{22}=0$ imposes the relation
\begin{equation}
  \label{eq:phi220}
  \dot M -3\frac{\ddot z}{\dot z}M=
\frac{\dot z^3}{2}
\frac{\partial (\eth_L \eth_L \bar\eth_L \bar\eth_L \,z
+ \bar\eth_L \bar\eth_L \eth_L \eth_L \,z
)}{\partial u} -
\dot z^2 \frac{\partial (\eth_L \eth_L z)}{\partial u} 
\frac{\partial (\bar\eth_L \bar\eth_L z)}{\partial u}.
\end{equation}
It is interesting to note that using $z$ the quantity $\mu$ acquires 
the compact form
\begin{equation}
  \label{eq:muz}
  \mu=\frac{i\dot z^3}{2}(
\bar\eth_L \bar\eth_L \eth_L \eth_L  -
\eth_L \eth_L \bar\eth_L \bar\eth_L )z.
\end{equation}

It is easy to see that this tetrad goes to the RT tetrad when $L$ goes to
zero.

The representation of the twisting algebraically special spacetimes in terms
of standard rotating null tetrads, we have just gave, is different from that
appearing in the book of exact solutions by Kramer et.al.\cite{Kramer80} and
its references, because we use a different tetrad frame.

The twist of the null congruence, and also the angular momentum of the
spacetime, are proportional to the quantity $c$ defined above.

The complex equation (\ref{eq:phi120}) and the real equation 
(\ref{eq:phi220}) constitute  three equations for the real functions $z$ and
$M$ and the complex function $L$; in other words one has three real
equations for four real scalars. In order to understand this under determined 
system it is important to mention the gauge freedom available.

\subsection{The gauge freedom}\label{sec:gaugeas}
These solutions are asymptotically flat.
The retarded time null coordinate $u$ defines, at future null infinity,
a set of sections by the equations $u=\tt constant$. The main gauge freedom can be
associated to a different choice of these sections; therefore let us consider
a transformation of the form
\begin{equation}
\tilde u = \tilde u_0(u,\zeta,\bar\zeta) + O(\frac{1}{r}) ,
\end{equation}
\begin{equation}
\tilde r = \frac{r}{w(u,\zeta,\bar\zeta)} + r_0(u,\zeta,\bar\zeta) ,
\end{equation}
\begin{equation}
\tilde \zeta = \zeta + O(\frac{1}{r}) ;
\end{equation}
one could also generalized the last equation to a transformation among the
sphere coordinates $(\zeta,\bar\zeta)$,
but they are not important for the following equations and therefore 
we do not 
consider them main gauge freedom.

The null character of the coordinate $\tilde u$ requires 
\begin{equation}
w = \dot{\tilde u}_0 .
\end{equation}

Although the vector $\tilde \ell$ is proportional to $\ell$, the other
tetrad vectors rotate under this transformation. From the study of the rotations
one deduces that
\begin{equation}\label{eq:tildev}
\tilde V = \frac{V}{w} ,
\end{equation}
\begin{equation}\label{eq:tildel}
\tilde V (\tilde L - \eth \tilde u_0) = V L
\end{equation}
and
\begin{equation}\label{eq:tildem}
\tilde M + i \tilde \mu = \frac{M+i\mu}{w^3} .
\end{equation}

Let us observe that there are several possibilities associated to the
three equations (\ref{eq:tildev}), (\ref{eq:tildel}) and
(\ref{eq:tildem}). If the twist of the null
congruence is different from zero then one can choose 
$u_0(u,\zeta,\bar\zeta)$ so that $\tilde V=1$, for example.

If we write the spin weight 1 quantity $L$ in terms of the 
spin weight 0 quantity $J$ by the relation
\begin{equation}\label{eq:J}
L=i\eth J = i \eth \Re e J - \eth \Im m J ;
\end{equation}
where the complex scalar $J$ can be expressed in term of its real and
imaginary parts $\Re e J$ and $\Im m J$, so that 
$J=\Re e J + i\Im m J$; then equation (\ref{eq:tildel}) can be written in
the form
\begin{equation}
i \eth_{\tilde V}\tilde J - \eth_{\tilde V} \tilde u_0 = i \eth_V J ;
\end{equation}
where $\eth_{\tilde V}$ is the edth operator for the sphere with metric
$dS^{2}=\frac{1}{\tilde V^2\,P_0^{2}}\;d\zeta \;d\bar{\zeta}$.
From this equation it can be seen that in principle one could use $u_0$
to make $\Im m \tilde J$ vanish; but it is not clear that the
needed $u_0$ would satisfy that $\dot u_0 > 0$; as it is required for 
a non-singular transformation. If $\Im m\dot J$ where zero then one can
see that a transformation of the form $\tilde u = u - \gamma(\zeta,\bar\zeta)$
would do the job. It is conceivable that for small time variations
one can use techniques similar to those of references 
\cite{Moreschi88} and \cite{Moreschi98}
to prove local existence of transformations that make $\Im m \tilde J = 0$.

Another possibility is to use the freedom involved in the choice of $u_0$ to make
$\tilde M$ to be a constant. This can always be made since 
from equation (\ref{eq:tildem}) one always solve for $\tilde u_0$, 
since
the function
$M$, for physically realistic spacetimes, is assumed to be positive definite.
This choice shows that from the four original unspecified real scalars, one
of them is pure gauge; therefore 
the complex equation
(\ref{eq:phi120}) and the real equation 
(\ref{eq:phi220})
constitute
a system of three equations for three physical real scalars.

\subsection{Algebraically special perturbations of RT spacetimes}\label{sec:asrtpert}

Having just presented the exact algebraically special solutions in terms of 
a new frame,
we proceed  here with our program by calculating the equations governing the 
algebraically special perturbations of RT
spacetimes. 
We will also specialize the discussion to the asymptotic
future regime for the retarded time going to infinity,
where the equations are solved.
 
When one studies the case of a linear perturbation of the RT line element, 
one considers:
\begin{equation}
V=V_{RT}+V_{L} 
\end{equation}
and
\begin{equation}
M=M_0 + M_L
\end{equation}
where $V_{RT}$ satisfies the RT equation, $M_0$ is the constant RT mass
 and  $V_{L}$ and $M_L$ represents the
departure from the RT metric that take into account the presence of angular momentum;
which are zero when $L$ vanishes.

From equation $\Phi_{12}=0$ it is obtained, in first order
\begin{equation} 
\begin{split} 
0=&
 -3 \,\dot L  M_0-\eth^3 \bar L \, \bar\eth\,V_{RT}\:  V_{RT}^{3}
 -3 \, \eth^2 \bar L \; \eth\,V_{RT}\: \bar\eth\,V_{RT}\:  V_{RT}^{2} 
 -3 \, \eth^2 \bar L\;  \bar\eth \eth\,V_{RT} \;  V_{RT}^{3} \\
& -\eth^2 V_{RT} \; \bar\eth \eth\,\bar L \;  V_{RT}^{3} 
 +\eth^2 V_{RT} \; \bar\eth^2 L \; V_{RT}^{3}
 -3 \,\eth\,L\: \bar\eth\,V_{RT}\:  V_{RT}^{3} \\
&-6 \,\eth \bar L\: \eth V_{RT}\: \bar\eth \eth V_{RT}\;  V_{RT}^{2}  
 -2 \,\eth\,\bar L\: \bar\eth \eth^2 V_{RT}\;  V_{RT}^{3} -\eth\,M_{L_1}
 -3 \,\eth^{2}\,V_{RT}\: \bar\eth \eth \bar L \;  V_{RT}^{2} \\
&+3 \eth^{2} V_{RT}\: \bar\eth^2 L\;  V_{RT}^{2} 
 -3 \,\eth V_{RT}\: \bar\eth \eth^2 \bar L\;  V_{RT}^{3} 
 +3 \,\eth V_{RT}\: \bar\eth \eth L\; \bar\eth V_{RT}\:  V_{RT}^{2} \\
&+6 \,\eth V_{RT}\: \bar\eth \eth V_{RT}\, \bar\eth L\:  V_{RT}^{2} 
 +3 \,\eth V_{RT}\: \bar\eth^2 \eth L\, V_{RT}^{3}
 -3 \,\eth V_{RT}\: \bar\eth L\:  V_{RT}^{3} \\
&-3 \,\eth V_{RT}\: \bar\eth V_{RT}\:  L \,V_{RT}^{2} 
 -\frac{1}{2} \bar\eth \eth^3 \bar L\;  V_{RT}^{4} 
 +\bar\eth \eth^2 L\, \bar\eth\,V_{RT}\:  V_{RT}^{3} \\
&+2 \,\bar\eth \eth^2 V_{RT}\, \bar\eth\,L\:  V_{RT}^{3} 
 +3 \,\bar\eth \eth L\, \bar\eth \eth V_{RT}\;  V_{RT}^{3} 
  -\frac{3}{2} \bar\eth \eth L\;  V_{RT}^{4} \\
&-3 \,\bar\eth \eth V_{RT}\;  L \;V_{RT}^{3} 
 +\frac{1}{2} \bar\eth^2 \eth^2 L\;  V_{RT}^{4}
\end{split} 
\end{equation} 
while from the equation $\Phi_{22}=0$ the following relation is obtained
\begin{equation} 
\begin{split} 
0=
&-2 \,\dot L \,\eth V_{RT}\: \bar\eth^2 V_{RT}\; V_{RT}^{3} 
 +2 \,\dot L \,\bar\eth^2 \eth V_{RT}\; V_{RT}^{4}
 + \dot L \,\bar\eth V_{RT}\: V_{RT}^{4}
 -2 \,\dot{\bar L} \,\eth^2 V_{RT}\, \bar\eth V_{RT}\: V_{RT}^{3} \\
&+2 \,\dot{\bar L} \,\bar\eth \eth^2 V_{RT}\; V_{RT}^{4}
 -\dot M_{L} \,V_{RT} 
 +3 \,\dot V_{L} \, M_0
 - \dot V_{RT}\, \eth^2 V_{RT}\, \bar\eth \bar L\: V_{RT}^{3} \\
&-\dot V_{RT}\;\eth L\: \bar\eth^2 V_{RT}\, V_{RT}^{3}
 +\frac{1}{2} \dot V_{RT}\; \eth \bar L\: V_{RT}^{4} 
 +\frac{1}{2} \dot V_{RT}\;\bar\eth \eth^2 \bar L\; V_{RT}^{4}
 +\frac{1}{2} \dot V_{RT}\;\bar\eth^2 \eth L\; V_{RT}^{4} \\
&-12 \,\frac{\dot V_{RT}}{V_{RT}} \,M_0 \,V_{L} 
 +3 \,\dot V_{RT}\, M_{L}
 +2 \, \eth \dot{\bar L}\, \bar\eth \eth V_{RT}\, V_{RT}^{4} 
 +\frac{1}{2} \eth \dot{\bar L}\, V_{RT}^{5}
 +\eth \dot V_{RT} \, \bar\eth \eth \bar L\, V_{RT}^{4} \\
&+\eth \dot V_{RT}\,\bar\eth^2 L\; V_{RT}^{4}
 -2\, \eth \dot V_{RT}\, \bar\eth^2 V_{RT}\; L \;V_{RT}^{3} 
 +\eth^2 \dot{\bar L}\; \bar\eth V_{RT}\: V_{RT}^{4} 
 +\eth^2 \dot V_{RT} \;\bar\eth \bar L\: V_{RT}^{4} \\
&+\eth^2 \bar L \; \bar\eth \dot V_{RT}\; V_{RT}^{4} 
 -\eth^2 V_{L}\; \bar\eth^2 V_{RT}\; V_{RT}^{3} 
 -2 \, \eth^2 V_{RT}\,\bar\eth \dot V_{RT}\, \bar L \; V_{RT}^{3}\\
&-\eth^2 V_{RT}\; \bar\eth^2 V_{L}\; V_{RT}^{3} 
 +\eth^2 V_{RT}\; \bar\eth^2 V_{RT}\; V_{L} \;V_{RT}^{2}
 +\eth\,L\: \bar\eth^2 \dot V_{RT}\; V_{RT}^{4} \\
&+2\, \eth \bar L\: \bar\eth \eth \dot V_{RT}\; V_{RT}^{4}
 +\eth V_{RT}\: \bar\eth \eth \dot{\bar L}\; V_{RT}^{4} 
 +\eth V_{RT}\: \bar\eth^2 \dot L \; V_{RT}^{4}
 +2\, \bar\eth \dot L \; \bar\eth \eth V_{RT}\; V_{RT}^{4}\\
&+\bar\eth \dot V_{RT}\;\bar\eth \eth L\; V_{RT}^{4} 
 +\bar\eth \dot V_{RT}\; L\; V_{RT}^{4}
 +\bar\eth \eth \dot L \;\bar\eth V_{RT}\: V_{RT}^{4} 
 +2 \,\bar\eth \eth \dot V_{RT}\;\bar\eth L\: V_{RT}^{4} \\
&+\frac{1}{2} \bar\eth \eth^2 \dot{\bar L}\; V_{RT}^{5}
 +2 \,\bar\eth \eth^2 \dot V_{RT}\;\bar L \;V_{RT}^{4} 
 +\frac{1}{2} \bar\eth^2 \eth \dot L \; V_{RT}^{5}
 +2\, \bar\eth^2 \eth \dot V_{RT}\; L \, V_{RT}^{4} \\
&+\bar\eth^2 \eth^2 V_{L}\; V_{RT}^{4}
\end{split} 
\end{equation} 

In the asymptotic future; for $u \rightarrow \infty$ it is 
known\cite{Frittelli92} that 
\begin{equation}
V_{RT} = 1 + O(e^{-\frac{2u}{M_0}});
\end{equation}
therefore to first order in the asymptotic expansion
of the retarded time, the above two equations become
\begin{equation}\label{eq:phi12c}
0=
-3 \dot L_1 \, M_0 - \eth\,M_{L_1}
-\frac{1}{2} \bar\eth \eth^3 \bar L_1  
-\frac{3}{2} \bar\eth \eth L_1  
+\frac{1}{2} \bar\eth^2 \eth^2 L_1  
;
\end{equation}
and
\begin{equation}\label{eq:phi22c}
0=
-\dot M_{L_1}
+3 \,\dot V_{L_1}\; M_0
+\frac{1}{2} \eth \dot{\bar L_1} 
+\frac{1}{2} \bar\eth \eth^2 \dot{\bar L_1}
+\frac{1}{2} \bar\eth^2 \eth \dot L_1 
+\bar\eth^2  \eth^2 V_{L_1}
;
\end{equation}
where the subindex 1 means that we are considering first order quantities in
the expansion as  $u \rightarrow \infty$.

Using the complex scalar $J$ which satisfies (\ref{eq:J}),
equations (\ref{eq:phi12c}) and (\ref{eq:phi22c}) become
\begin{equation}\label{eq:phi12d}
0=
3 \eth\left(\Im m \dot J_1 \right) i M_0
+3 \eth\left(\Re e \dot J_1 \right) M_0
-\eth\,M_{L_1}\: i
-\eth(\bar\eth^2 \eth^2 \Re e J_1  ) 
\end{equation}
and
\begin{equation}\label{eq:phi22d}
0=
-\dot M_{L_1}
+3\, \dot V_{L_1} \; M_0
-\bar\eth^2 \eth^2 \Im m \dot J_1 
+\bar\eth^2 \eth^2 V_{L_1}
;
\end{equation}
where it can be seen from equation (\ref{eq:phi12d}) that if we use the gauge
freedom to make $M_{L_1}$ a constant, then $\eth \Im m \dot J_1$ must vanish
which implies that $\Im m \dot J_1=0$. 
It is clear from the discussion of the previous subsection that by means
of a further gauge transformation one can make  $\Im m J_1=0$.
Therefore in this gauge the above equations
become
\begin{equation}\label{eq:phi12e}
0 = \eth \left( 3\,   \dot J_1 \,M_0 
-  \bar\eth^2 \eth^2  J_1 \right)
\end{equation}
where now $J_1$ is assumed to be real, and
\begin{equation}\label{eq:phi22e}
0 =
  3\dot V_{L_1} M_0
+ \bar\eth^2 \eth^2 V_{L_1}  .
\end{equation}

The constant $M_{L_1}$ now plays no role, and it can be made to vanish by a further
reparameterization of the retarded time coordinate. Integrating equation 
(\ref{eq:phi12e})
we can write these two equations in a more comparable form
\begin{equation}\label{eq:phi12f}
 3\,   \dot J_1 \,M_0 =  \bar\eth^2 \eth^2  J_1 
\end{equation}
and
\begin{equation}\label{eq:phi22f}
- 3\dot V_{L_1} \,M_0 = \bar\eth^2 \eth^2 V_{L_1}  .
\end{equation}

It is important to know the spectrum of the operator appearing on the right hand
side. The eigenvalues of the operator 
$\bar\eth^2 \eth^2$ 
are\cite{Moreschi88}
$\frac{(\ell-1)\ell(\ell+1)(\ell+2)}{4}$; where $\ell$ is any non-negative
integer.

Therefore the leading order behavior for the functions are:
\begin{equation}
J_1 = J_1^m \,Y_{1m}(\zeta,\bar\zeta) 
      +J_2^m \,Y_{2m}(\zeta,\bar\zeta) e^{\frac{2 u}{M_0}} + O(e^{\frac{4 u}{M_0}})
\end{equation}
and
\begin{equation}
V_{L_1} = V_{L_1}^m\,Y_{2m}(\zeta,\bar\zeta) 
e^{-\frac{2 u}{M_0}} + O(e^{-\frac{4 u}{M_0}});
\end{equation}
where $J_1^m$, $J_2^m$ and $V_{L_1}^m$ are constants and we have used 
gauge transformations to set to zero spurious integrations constants. 

It is observed that $V_{L_1}$
has the same asymptotic behavior as $V_{RT}$; while $J_1$ has a 
constant angular momentum term and a possible
divergent behavior for the higher moments in the limit
$u \rightarrow \infty$.

Therefore these spacetimes are not
suitable for the description of a perturbed black hole with a
dynamic angular momentum. This forces us to
consider general perturbations of RT spacetimes, that we treat in the
next section.

\section{General perturbations of RT spacetimes}\label{sec:general}

\subsection{The vacuum equations}

Let us consider now a general perturbation of the RT spacetimes.
We will still work with a null tetrad satisfying equations 
(\ref{eq:produc}-\ref{eq:vecn}), 
with $\xi^0=0$, and the leading order of $\xi^2$ and $\xi^3$
also given by eq. (\ref{eq:xileading}); but where now the scalar
$V$ is given by
\begin{equation}
  \label{eq:vpertur}
  V=V_{RT} + \lambda \, V_\lambda ;
\end{equation}
where  $V_{RT}$ is the RT scalar  discussed in section \ref{sec:RT}, 
$\lambda$ the linearization parameter
and $V_\lambda$ the linear perturbation scalar. 
Let us emphasize that this means that the tetrad vector $\ell^a$
has no twist, unlike the corresponding one in
the rotating null tetrad used in the last section.
In this section we use asymptotic coordinates and standard tetrad
as defined in \cite{Perez01}; however it is important to notice that
we do not use a Bondi system.
We use $\lambda$ to emphasize in some  equations the linearized terms;
but it should be understood that it has the value 1 at any stage.

It is convenient to start the study from the vacuum Bianchi identities.

Assume that in the null hypersurface $\Sigma$, defined by $u=u_0$, 
of the original RT spacetime one is given the function 
$\Psi_0(u_0,r,\zeta,\bar\zeta)$.
Then, noting that $\Psi_0$ and $\Psi_1$ are zero in the Robinson-Trautman
geometries; 
one obtains, from eq. (3.91) in ref. \cite{Moreschi87} that, on $\Sigma$,
$\Psi_1$ is given by
\begin{equation}
  \label{eq:Psi1r}
  \Psi_1 = \frac{\Psi_1^0}{r^4}+
\frac{1}{r^4}\int^r {r'}^3 \, \bar\eth_{V_{RT}} \, \Psi_0 \, dr' ;
\end{equation}
where $\Psi_1^0=\Psi_1^0(u_0,\zeta,\bar\zeta)$.

Similarly, from eq, (3.90) of  ref. \cite{Moreschi87} one obtains that
$\Psi_2$ is given  on $\Sigma$ by
\begin{equation}
  \label{eq:Psi2r}
  \Psi_2 = \frac{\Psi_2^0}{r^3}+
\frac{1}{r^3}\int^r {r'}^2 \, \bar\eth_{V_{RT}} \, \Psi_1 \, dr' ;
\end{equation}
with
\begin{equation}
\Psi_2^0= -\left( M_0+ \lambda (M_1 + i \, \mu) \right),
\end{equation} 
where $M_0$ is the constant RT mass,  
$M_1=M_1(u_0,\zeta,\bar\zeta)$ and $\mu=\mu(u_0,\zeta,\bar\zeta)$ are
real functions and $\eth_{V_{RT}}$ is the edth operator of the
2-sphere with the RT conformal factor. 

When calculating $\Psi_3$ from eq. (3.89) of ref. \cite{Moreschi87}, one 
should take into account the first order contribution coming from the
edth primed operator; due to the fact that $\Psi_2$ has zero order terms.

In order to proceed, we need to calculate the first order perturbation
of the spin coefficients $\rho$ and $\sigma$. 

Equation (3.50) of \cite{Moreschi87} is
\begin{equation}
  \label{eq:rhor}
  \frac{\partial \rho}{\partial r}=\rho^2 ;
\end{equation}
which has solution
\begin{equation}
  \label{eq:rhopert}
  \rho = -\frac{1}{r} ;
\end{equation}
after appropriately chosen the freedom in the origin of the coordinate $r$.

Then eq. (3.51) of \cite{Moreschi87} becomes
\begin{equation}
  \label{eq:sigmar}
    \frac{\partial \sigma}{\partial r}= 2 \rho \sigma + \Psi_0 ;
\end{equation}
which has solution
\begin{equation}
  \label{eq:sigmapert}
  \sigma = \frac{\sigma_0}{r^2}+ \frac{1}{r^2}\int^r {r'}^2 \, \Psi_0 \,dr' ;
\end{equation}
where $\sigma_0=\sigma_0(u_0,\zeta,\bar\zeta)$.

From eq. (3.52) of \cite{Moreschi87} one obtains
\begin{equation}
  \label{eq:taur}
    \frac{\partial \tau}{\partial r}= 2 \rho \tau + \Psi_1 ;
\end{equation}
with  solution
\begin{equation}
  \label{eq:taupert}
  \tau = \frac{\tau_0}{r^2}+ \frac{1}{r^2}\int^r {r'}^2 \, \Psi_1 \,dr' ;
\end{equation}
where $\tau_0=\tau_0(u_0,\zeta,\bar\zeta)$.

One can now integrate eq. (3.31) of \cite{Moreschi87} to obtain the
radial dependence of $\xi^i$. It is deduced that
\begin{equation}
  \label{eq:xir1}
  \xi^i = \frac{\xi^i_0}{r} + 
          \frac{1}{r} \int^r r'\,  \sigma \,\bar\xi^i \, dr';
\end{equation}
where $\xi^i_0$ does not depend on $r$; then,
one can write the integral equation
\begin{equation}
  \label{eq:xiint}
  \xi^i = \frac{\xi^i_0}{r} + 
          \frac{1}{r} \int^r r'\,  \sigma \,
 \left( \frac{\bar\xi^i_0}{r'} + 
          \frac{1}{r'} \int^{r'} r''\,  \bar\sigma \, \xi^i \, dr'' \right)
 \, dr';
\end{equation}
which can be solved by iterations. But, since we should consider only linear
terms on $\lambda$, one obtains that in first order
\begin{equation}
  \label{eq:xipert}
\begin{split}
  \xi^i &= \frac{\xi^i_0}{r} + 
          \frac{1}{r} \int^r   \sigma \, \bar\xi^i_0  \, dr' \\
&=  \frac{\xi^i_0}{r} 
- \frac{ \bar\xi^i_0 \,\sigma_0 }{{r}^2}
+ 
          \frac{ \bar\xi^i_0 }{r} \int^r  
\left(
\frac{1}{{r'}^2}\int^{r'} {r''}^2 \, \Psi_0 \,dr'' 
\right)
\,  dr' 
.
\end{split}
\end{equation}

The radial dependence of the components $X^i$ is obtained from eq. (3.28) of
\cite{Moreschi87}; which is
\begin{equation}
  \label{eq:xr}
  \frac{\partial X^i}{\partial r}= 2\bar\tau \, \xi^i + 2\tau\, \bar\xi^i ;
\end{equation}
which implies, in first order, that
\begin{equation}
\begin{split}
  \label{eq:xpert}
  X^i &= X^i_0 + \left[ 2 \bar\xi^i_0
\int^r 
\left(
 \frac{\tau_0}{{r'}^3}+ \frac{1}{{r'}^3}\int^{r'} {r''}^2 \, \Psi_1 \,dr''
\right) 
dr' + {\tt c.c.} \right]
;
\end{split}
\end{equation}
where {\tt c.c.} means complex conjugate and the $X^i_0 $ do not depend 
on $r$; but from the coordinate condition at
future null infinity\cite{Perez01}, one actually  has   $X^i_0=0$.

In order to determine the dependence of $U$ on the coordinate $r$, 
one could use
eq. (3.26) of \cite{Moreschi87}; which requires to calculate $\epsilon'$ 
first. Equation (3.58) of the same reference is in our case
\begin{equation}
  \label{eq:epspr}
  \frac{\partial \epsilon'}{\partial r} =
- 2 \bar\tau \, \beta + 2 \tau \, \beta' - \Psi_2;
\end{equation}
therefore
\begin{equation}
  \label{eq:epspr2}
\begin{split}
  \frac{\partial (\epsilon'+\bar\epsilon')}{\partial r} &=
- 2 \bar\tau \, (\beta - \bar\beta')+ 2 \tau \, (\beta' - \bar\beta) 
- (\Psi_2+\bar\Psi_2) \\
&= -4 \tau \bar\tau - (\Psi_2+\bar\Psi_2)
;
\end{split}
\end{equation}
where we must neglect the product involving $\tau$; since it is a quantity of
higher order. One concludes that
\begin{equation}
  \label{eq:urr}
  \frac{\partial^2 U}{\partial r^2}= - (\Psi_2+\bar\Psi_2);
\end{equation}
which implies that $U$ is of the form
\begin{equation}
  \label{eq:upert}
  U=r\, U_{00} + U_0 - \int^r \int^{r'} (\Psi_2 + \bar\Psi_2) dr'' \, dr';
\end{equation}
where $U_{00}$ and $U_0$ do not depend on $r$.

The $r$ dependence of the tetrad components can be explicitly given in terms of
the potential $W_0(u,r,\zeta,\bar\zeta)$, defined from
\begin{equation}\label{eq:poten}
\Psi_0 = \frac{\partial^4 W_0}{\partial r^4} ,
\end{equation}
by the relations
\begin{equation}
\begin{split}
\xi ^{0} & =0, \\
\xi ^{2} & =\frac{\xi _{0}^{2}}{r} + \lambda \bar{\xi}_{0}^{2} 
\left(- \frac{\sigma_{0}}{r^{2}}
  + \frac{1}{r}\frac{\partial^2 W_0}{\partial r^2}
  - \frac{2}{r^{2}} \frac{\partial W_0}{\partial r}
\right)
, \\
\xi ^{3} & =\frac{\xi _{0}^{3}}{r} + \lambda \bar{\xi}_{0}^{3} 
\left(- \frac{\sigma_{0}}{r^{2}}
  + \frac{1}{r}\frac{\partial^2 W_0}{\partial r^2}
  - \frac{2}{r^{2}} \frac{\partial W_0}{\partial r}
\right)
\end{split}
\end{equation}
with
\begin{equation}
\xi _{0}^{2}=\sqrt{2}P_{0}\;V,\quad \xi _{0}^{3}=-i\xi _{0}^{2} ;
\end{equation}
\begin{equation}
U=rU_{00}+U_{0}+\frac{U_{1}}{r}+\frac{U_{2}}{r^{2}} +\Delta U_{3} ,
\end{equation}
where
\begin{equation}
\begin{split}
U_{00} & =\frac{\dot{V}}{V}, \\
U_{0} & =-\frac{1}{2}K_{V}, \\
U_{1} & =-\frac{1}{2}\left(\Psi _{2}^{0}+\bar{\Psi}_{2}^{0}\right) ,\\ 
U_{2} & =\frac{\lambda}{6} \left( \eth_{V_{RT}}\bar{\Psi}_{1}^{0}
             +\bar{\eth }_{V_{RT}}\Psi _{1}^{0}\right) , \\
\Delta U_3 & = - \frac{\lambda}{r^2} \left( \bar\eth_{V_{RT}}^2 W_0 
  + \eth_{V_{RT}}^2 \bar W_0 \right);
\end{split}
\end{equation}
and
\begin{equation}
\begin{split}
X^{0} & =1, \\
X^{2} & =\lambda \xi _{0}^{2}
          \left( -  \frac{\bar{\tau}_{0}}{r^{2}}
                 + \frac{2 \bar{\Psi}_{1}^{0}}{3r^{3}}
                 + \frac{2}{r^2}\frac{\partial \eth_{V_{RT}} \bar W_0}
                                     {\partial r}
                 - \frac{4}{r^3}\eth_{V_{RT}} \bar W_0
          \right) + {\tt c.c.} , \\
X^{3} & =\lambda \xi _{0}^{3}
          \left( -  \frac{\bar{\tau}_{0}}{r^{2}}
                 + \frac{2 \bar{\Psi}_{1}^{0}}{3r^{3}}
                 + \frac{2}{r^2}\frac{\partial \eth_{V_{RT}} \bar W_0}
                                     {\partial r}
                 - \frac{4}{r^3}\eth_{V_{RT}} \bar W_0
          \right) + {\tt c.c.} ;
\end{split}
\end{equation}
where
\begin{equation}
\tau_0 = \bar\eth_{V_{RT}}\sigma_0,
\end{equation}
\begin{equation}
K_{V}=\frac{2}{V}\bar{\eth }_{V}\eth _{V}V-\frac{2}{V^{2}}\eth _{V}V~%
\bar{\eth }_{V}V+V^{2}.
\end{equation}
In these equations we are explicitly denoting the first order terms by
introducing the $\lambda$ dependency.
The scalar $V$ is given by equation (\ref{eq:vpertur}), and therefore one can 
express $K_V$ by
\begin{equation}
K_{V}=K_{V_{RT}}+\lambda ~K_{V_{\lambda }},
\end{equation}
where
\begin{equation}
K_{V_{RT}}=\frac{2}{V_{RT}}\bar{\eth }_{V_{RT}}\eth _{V_{RT}}V_{RT}
-\frac{2}{V_{RT}^{2}}\eth _{V_{RT}}V_{RT}~\bar{\eth }_{V_{RT}}V_{RT}
+V_{RT}^{2},
\end{equation}
and
\begin{multline}
K_{V_\lambda}=\frac{2}{V_{RT}}\bar{\eth }_{V_{RT}}\eth _{V_{RT}}V_\lambda
-\frac{2}{V_{RT}^{2}}\eth _{V_{RT}}V_\lambda~\bar{\eth }_{V_{RT}}V_{RT} 
-\frac{2}{V_{RT}^{2}}\eth _{V_{RT}}V_{RT}~\bar{\eth }_{V_{RT}}V_\lambda \\
+\frac{2V_\lambda}{V_{RT}^{3}}\eth _{V_{RT}}V_{RT}
         ~\bar{\eth }_{V_{RT}}V_{RT} 
+ V_\lambda \, V_{RT} + \frac{V_\lambda}{V_{RT}} \, K_{V_{RT}}
.
\end{multline}

The previous relations make the following spinor components of 
the Ricci tensor  to vanish
\begin{equation}
\Phi_{00}=0 ,
\end{equation}
\begin{equation}
\Phi_{01}=0 ,
\end{equation}
\begin{equation}
\Phi_{11} =0
\end{equation}
and
\begin{equation}
\Lambda =0 .
\end{equation}

The other equations, coming from the other spinor components, involve 
time derivatives of the functions. They are:
\begin{multline}\label{eq:w0u}
\frac{\partial^2 \left(\frac{W_0}{r^3}\right)}{\partial u \partial r}=
 - \frac{1}{r^2}\frac{\dot V_{RT}}{V_{RT}}\frac{\partial^2 W_0}{\partial r^2}
+ \frac{1}{r^3}\left( 4 \frac{\dot V_{RT}}{V_{RT}}\frac{\partial W_0}
  {\partial r} + \frac{1}{2} K_{V_{RT}} \frac{\partial^2 W_0}{\partial r^2}\right)\\
+ \frac{1}{r^4}\left( -6 \frac{\dot V_{RT}}{V_{RT}} W_0
  -  K_{V_{RT}} \frac{\partial W_0}{\partial r}
   - M_0 \frac{\partial^2 W_0}{\partial r^2} \right) \\
+ \frac{1}{r^5}\left( -\frac{1}{6} \eth_{V_{RT}}\Psi_1^0 
   + \bar\eth_{V_{RT}}\eth_{V_{RT}} W_0
   + M_0 \frac{\partial W_0}{\partial r}
   + \frac{1}{2} M_0 \, \sigma_0 
   - 2 K_{V_{RT}}  W_0  \right),
\end{multline}

\begin{equation}\label{eq:psi1u}
\dot \Psi_1^0=
 3\frac{\dot V_{RT}}{V_{RT}} \Psi_1^0 - \eth_{V_{RT}} (M_1 + i\mu)
 + \bar\eth_{V_{RT}}\left(K_{V_{RT}}\right)\sigma_0,
\end{equation}
where $\mu$ is actually related to $\sigma_0$ by
\begin{equation}\label{eq:mu}
\mu = \frac{1}{2i}\left(\eth_{V_{RT}}^2 \bar\sigma_0 -
                    \bar\eth_{V_{RT}}^2 \sigma_0 \right)
;
\end{equation}
the zero order term  of $\Phi_{22}= 0$ gives
\begin{equation}\label{eq:rt2}
-6M_0 \frac{\dot{V}_{RT}}{V_{RT}}=\;\bar{\eth }_{V_{RT}}\eth
_{V_{RT}}\,K_{V_{RT}}, 
\end{equation}
which is just another way to write the RT equation, 
and from the first order one obtains 
\begin{multline}\label{eq:vla}
-6M_0 \frac{\dot{V}_{\lambda }}{V_{RT}} =
\;\bar{\eth }_{V_{RT}}\eth _{V_{RT}}~K_{V_{\lambda }}
- 6\frac{\dot{V}_{RT}}{V_{RT}}\left(3M \frac{V_\lambda}{V_{RT}} - M_{1}\right)
- 2\dot{M}_{1} \\ 
-   \frac{\dot{V}_{RT}}{V_{RT}} \bar\eth_{V_{RT}}^2 \sigma_0 
+ 2 \frac{\dot{V}_{RT}}{V_{RT}^2} \bar\eth_{V_{RT}} V_{RT} 
                                  \bar\eth_{V_{RT}}\sigma_0 
- \frac{2}{V_{RT}}  \bar\eth_{V_{RT}} \dot V_{RT} 
                                  \bar\eth_{V_{RT}}\sigma_0 
+  \bar\eth_{V_{RT}}^2 \dot{\sigma}_0 \\
-   \frac{\dot{V}_{RT}}{V_{RT}} \eth_{V_{RT}}^2 \bar\sigma_0 
+ 2 \frac{\dot{V}_{RT}}{V_{RT}^2} \eth_{V_{RT}} V_{RT} 
                                  \eth_{V_{RT}}\bar\sigma_0 
- \frac{2}{V_{RT}}  \eth_{V_{RT}} \dot V_{RT} 
                                  \eth_{V_{RT}}\bar\sigma_0
+ \eth_{V_{RT}}^2 \dot{\bar\sigma}_0 
.
\end{multline}

One could also consider the leading order asymptotic expansion of $W_0$ by writing
\begin{equation}
W_0 =  \frac{\Psi_0^0}{4!r} + W_1 ;
\end{equation}
where $\Psi_0^0=\Psi_0^0(u,\zeta.\bar\zeta)$;
in this way one has
\begin{equation}
\Psi_0=  \frac{\Psi_0^0}{r^5} + \frac{\partial^4 W_1}{\partial r^4},
\end{equation}
where the second term is of order $O(1/r^6)$.

In this case equation (\ref{eq:w0u}) splits into two equations, namely
\begin{equation}\label{eq:psi0u}
\dot\Psi_0^0 = 3\frac{\dot V_{RT}}{V_{RT}}\Psi_0^0 + \eth_{V_{RT}}\Psi_1^0
 -3M_0 \,\sigma_0
\end{equation}
and
\begin{multline}\label{eq:w1u}
\frac{\partial^2 \left(\frac{W_1}{r^3}\right)}{\partial u \partial r}=
 - \frac{1}{r^2}\frac{\dot V_{RT}}{V_{RT}}\frac{\partial^2 W_1}{\partial r^2}
+ \frac{1}{r^3}\left( 4 \frac{\dot V_{RT}}{V_{RT}}\frac{\partial W_1}
  {\partial r} + \frac{1}{2} K_{V_{RT}} \frac{\partial^2 W_1}
{\partial r^2}\right)\\
+ \frac{1}{r^4}\left( -6 \frac{\dot V_{RT}}{V_{RT}} W_1
  -  K_{V_{RT}} \frac{\partial W_1}{\partial r}
   - M_0 \frac{\partial^2 W_1}{\partial r^2} \right) \\
+ \frac{1}{r^5}\left( 
   + \bar\eth_{V_{RT}}\eth_{V_{RT}} W_1
   + M_0 \frac{\partial W_1}{\partial r}
   + \frac{1}{2} M_0 \, \sigma_0  - 2 K_{V_{RT}}  W_1 \right) \\
+ \frac{1}{r^6}\left(  
   + \frac{1}{24}\bar\eth_{V_{RT}}\eth_{V_{RT}} \Psi_0^0
   \right)
- \frac{1}{8r^7}\Psi_0^0 M_0  
.
\end{multline}

Let us notice that equations (\ref{eq:w1u}), (\ref{eq:psi0u}), (\ref{eq:psi1u})
and (\ref{eq:vla}) are the evolution equations for the first order
quantities $W_1$, $\Psi_0^0$, $\Psi_1^0$ and $V_\lambda$ and $M_1$ 
respectively; while equation (\ref{eq:rt2}) 
(the RT equation)
is the evolution equation
for the zero order quantity $V_{RT}$. The scalar $\sigma_0$ is
an arbitrary complex function that affects the radiation content
of the spacetime. To understand these equations it is important
to study the gauge freedoms considered in the next subsection.

\subsection{The gauge freedom}

The main gauge freedom admitted in our calculation is of the form
\begin{align}
\tilde u &= \tilde u_0(u,\zeta,\bar\zeta) + 
 \frac{\tilde u_1(u,\zeta,\bar\zeta)}{r} 
+ O\left(\frac{1}{r^2}\right) ,\\
\tilde r &= \frac{r}{w(u,\zeta,\bar\zeta)} 
+ O\left(r^0\right)  ,\\
\tilde \zeta & = \zeta + O\left(\frac{1}{r}\right) ;
\end{align}
one could also admit a further transformation of the coordinates of the sphere
$(\zeta,\bar\zeta)$ into itself, that was also available in the original 
Robinson-Trautman geometry; but this will not change the following result.

The condition $g^{\tilde u \tilde r}=1$ imposes the relation
\begin{equation}
w = \dot{\tilde u}_0 .
\end{equation}

This asymptotic coordinate transformation is associated to a corresponding
null tetrad transformation; which in the leading orders is given by
\begin{equation}
\begin{split}
\tilde \ell & = d\tilde u = \dot{\tilde u}_0 \, du
+ \tilde u_{0\zeta}  \, d\zeta + \tilde u_{0\bar\zeta}  \, d\bar\zeta 
+O\left(\frac{1}{r}\right) \\
&=  \dot{\tilde u}_0  \, \ell
- \frac{\eth_{V} \tilde u_0}{r}  \, \bar m 
- \frac{\bar\eth_V \tilde u_0}{r}  \, m 
+O\left(\frac{1}{r}\right) .
\end{split}
\end{equation}

\begin{equation}
\begin{split}
\tilde n &= 
 \frac{\partial}{\partial \tilde u}  + O\left( \frac{1}{r}\right)
 =  \frac{1}{\dot{\tilde u}_0}  \,  \frac{\partial}{\partial u}
+ O\left( \frac{1}{r}\right)\\
&=  \frac{1}{\dot{\tilde u}_0}  \, n
+ O\left( \frac{1}{r}\right) ,
\end{split}
\end{equation}

\begin{equation}
\begin{split}
\tilde m &= 
\frac{\sqrt{2}\tilde P}{\tilde r}  \frac{\partial}{\partial \tilde \zeta}  
  + O\left( \frac{1}{r^2}\right)
 = \frac{\sqrt{2}P_0 \tilde V\, w }{ r}  
\left( -\frac{\tilde u_{0\zeta}}{\dot{\tilde u}_0} 
\,  \frac{\partial}{\partial u}
+ \frac{\partial}{\partial \zeta} \right)
+ O\left( \frac{1}{r^2}\right)\\
&=  -\frac{\sqrt{2}P_0 \tilde V\, w }{ r}  
\frac{\tilde u_{0\zeta}}{\dot{\tilde u}_0} \,  n
+ \frac{\tilde V \, w}{V}m
+ O\left( \frac{1}{r^2}\right) ;
\end{split}
\end{equation}
since the metric expressed in terms of the new null tetrad 
must coincide with the metric
expressed in terms of the original null tetrad, it is deduced that
\begin{equation}\label{eq:tildeV}
\tilde V = \frac{V}{w}=\frac{V}{\dot{\tilde u}_0} ;
\end{equation}
therefore
\begin{equation}
\tilde m = m
  -\frac{\eth_V \tilde u_{0\zeta}}{r \, \dot{\tilde u}_0 } \,  n
+ O\left( \frac{1}{r^2}\right) .
\end{equation}

The null tetrad transformation equations can be used to write the leading order
transformation relations for the spinor dyad associated to the 
null tetrad\cite{Geroch73}; namely
\begin{equation}
\tilde o^A = \sqrt{\dot{\tilde u}_0} 
\left( o^A - \frac{\eth_{V} \tilde u_0}{r\, \dot{\tilde u}_0}\, \iota^A \right),
\end{equation}
\begin{equation}
\tilde \iota^A = \frac{1}{\sqrt{\dot{\tilde u}_0}}\, \iota^A ;
\end{equation}
which implies that the regular dyad at future null infinity is given by
\begin{equation}
\hat{\tilde o}^A = \tilde\Omega^{-1}\, \tilde o^A = 
\frac{r}{w}\sqrt{\dot{\tilde u}_0} 
\left( o^A - \frac{\eth_{V} \tilde u_0}{r\, \dot{\tilde u}_0}\, \iota^A \right)
=\frac{1}{\sqrt{\dot{\tilde u}_0}}
\left(\hat o^A - \frac{\eth_{V} \tilde u_0}{\dot{\tilde u}_0}\, 
\hat\iota^A \right)
,
\end{equation}

\begin{equation}
\hat{\tilde \iota}^A = \tilde \iota^A = \frac{1}{\sqrt{\dot{\tilde u}_0}}\, 
\hat \iota^A .
\end{equation}
We can now easily calculate the component $\Psi_2$ of the Weyl tensor, 
in leading order,
with respect to the new null tetrad, obtaining
\begin{equation}
\tilde \Psi_2^0 = \tilde \Omega \Psi_{ABCD}
\hat{\tilde o}^A \hat{\tilde o}^B \hat{\tilde \iota}^C \hat{\tilde \iota}^D
= \frac{1}{\dot{\tilde u}_0^3} 
\left( \Psi_2^0 - \frac{2\, \eth_V\, \tilde u_0}{\dot{\tilde u}_0} \Psi_3^0
+ \frac{(\eth_V \tilde u_0)^2}{\dot{\tilde u}_0^2}\, \Psi_4^0 \right) ;
\end{equation}
where in our case we have
\begin{equation}
\Psi_3^0 =\eth_V {\sigma'}_0 - \bar\eth_V {\rho'}_0 
= -\frac{1}{2}\bar\eth_{V_{RT}}  K_{V_{RT}} + O(\lambda)
,
\end{equation}
and
\begin{equation}
\Psi_4^0 =\dot {\sigma'}_0 - 2\,U_{00}\, {\sigma'}_0 - \bar\eth_V^2 U_{00}
= \bar\eth_{V_{RT}}^2  \frac{\dot V_{RT}}{V_{RT}} +O(\lambda)
\end{equation}
where
\begin{equation}
{\sigma'}_0 =  \lambda \left( \frac{\dot V_{RT}}{V_{RT}} \bar\sigma_0
                              - \dot{ \bar\sigma_0} \right),
\end{equation}
and
\begin{equation}
{\rho'}_0 =  \frac{1}{2} K_V .
\end{equation}

Let us consider a first order transformation generated by
\begin{equation}
\tilde u_0 = u - \lambda \, \gamma(u,\zeta,\bar\zeta) ;
\end{equation}
then one has
\begin{equation}
\begin{split}
\tilde \Psi_2^0 &= 
- \left[
M_0 + \lambda (\tilde M_1 + i \tilde \mu)
\right] \\
& = \frac{1}{(1- \lambda \dot{\gamma})^3} 
\left( - \left[
M_0 + \lambda ( M_1 + i \mu)
\right]
- \lambda \, \eth_{V_{RT}}\gamma \,\,\bar\eth_{V_{RT}}  K_{V_{RT}}  
\right) ;
\end{split}
\end{equation}
where it is important to note that the transformation law for
$\sigma_0$ is
\begin{equation}
\lambda \tilde\sigma_0 = \lambda \, \sigma_0 - \lambda \, \eth_{V_{RT}}^2 \gamma
\end{equation}
and therefore, since $\gamma$ is real, it can be seen from 
equation (\ref{eq:mu}) that
\begin{equation}
\tilde \mu = \mu .
\end{equation}
Then,  one concludes that
\begin{equation}\label{eq:tildeM1}
\tilde M_1  = 
3M_0\, \dot\gamma +   M_1 
+ \, \eth_{V_{RT}}\gamma \,\,\bar\eth_{V_{RT}}  K_{V_{RT}}   ;
\end{equation}
where it is observed that the left hand side is real, and the first two terms of the
right hand side are also real; therefore for this equation to make sense one should
have the last term also real, which at first sight is not obvious. In order to check this 
let us note that for any quantity $H$ of spin weight $s$ one has
\begin{equation}
\eth_V \bar\eth_V H - \bar\eth_V \eth_V H = - s \, K_V\, H ;
\end{equation}
then, taking $H=\eth_V \gamma$ one has $s=1$ and
\begin{equation}
\eth_V \bar\eth_V \eth_V \gamma - \bar\eth_V \eth_V^2 \gamma  = - \, K_V\, \eth_V \gamma ;
\end{equation}
so that acting with $\bar\eth_V$ on the last expression one obtains
\begin{equation}
\bar\eth_V \eth_V \bar\eth_V \eth_V \gamma - \bar\eth_V^2 \eth_V^2 \gamma = 
- \bar\eth_V K_V\, \eth_V \gamma - K_V \bar\eth_V \eth_V \gamma .
\end{equation}
Since $ \bar\eth_V \eth_V$ and $ \bar\eth_V^2 \eth_V^2$ are real operators, one concludes that
$  \bar\eth_V K_V\, \eth_V \gamma$ is a real quantity for any $V$.

Coming back to equation (\ref{eq:tildeM1}) we observe that given $M_1(u,\zeta\bar\zeta)$ and
$V_{RT}(u,\zeta,\bar\zeta)$ it is possible to chose $\gamma(u,\zeta,\bar\zeta)$ such that
$\tilde M_1$ acquires any value that we want.

Alternatively, from equation (\ref{eq:tildeV}) we  have
\begin{equation}
\tilde V = V_{RT} + \lambda ( V_\lambda + \dot \gamma) ;
\end{equation}
therefore we could also choose $\gamma$ in order to give to $\tilde V_\lambda$ 
any desired value.

Summarizing, equations 
(\ref{eq:psi0u}), (\ref{eq:psi1u}) and (\ref{eq:rt2})
are the evolution equations for the asymptotic quantities $\Psi_0^0$, $\Psi_1^0$ 
and $V_{RT}$ respectively. And, equation (\ref{eq:vla}) can be understood as the
evolution equation for $V_\lambda$ if $M_1$ is known; alternatively,
it can be understood as the evolution equation of $M_1$ if $V_\lambda$ is given.
As it was mentioned previously, the function $\sigma_0$ is free data; 
which contributes to the description of the
structure of the sources and to outgoing radiation.

\section{Final comments}\label{sec:final}

In subsection \ref{sec:asexact} we have presented in full detail
the algebraically special vacuum solutions in a new frame. The gauge 
freedom of these solutions is discussed in subsection \ref{sec:gaugeas}.

The algebraically special perturbations of RT spacetimes,
studied in section \ref{sec:asrtpert}, are shown
to contain divergent behavior in the asymptotic future, and therefore
they do not seem to provide with an appropriate model 
for the description of a binary collision. This result contrasts with
the usual treatments of perturbations of black holes. For example
in the study of gravitational perturbations of the Kerr geometry,
one normally assumes perturbations which, with respect to the
Boyer Lindquist coordinates, have a $t$ and $\phi$ dependence given by
$e^{i(\hat \Sigma t + m \phi)}$\cite{Chandra79}; where $\hat \Sigma$ is a
constant considered ``mostly real and positive''\cite{Chandra92}. 
However it is interesting to note that equation (7.286) of
reference \cite{Chandra79} reads
\begin{equation*}
{\cal C} = D - 12 i \hat \Sigma M
\end{equation*}
where $D$ is a positive real constant, $M$ the mass of the spacetime
and if particularized to the case of algebraically special perturbation;
namely $\Psi_0=0$, then the complex constant $\cal C$ must be zero; 
which indicates that $e^{i(\hat \Sigma t + m \phi)}$ has an
exponential divergent behavior for late times, as we have found.

One could also ask, how is that this behavior does not appear in the
quasi-normal modes study of Schwarzschild black-hole; and the answer
is that the very definition of quasi-normal modes excludes
the exponentially growing  $\Im m \;\hat \Sigma <0$ cases\cite{Chandra92}.
It is probably worth while to remark that the quasi-normal modes 
can not describe the Robinson-Trautman geometry as a perturbation of the
Schwarzschild metric.

In relation to the general perturbations of the RT spacetimes,
we have shown in section \ref{sec:general} that they provide
with a suitable family of spacetimes for the discussion of the estimate of
the total energy radiated in the collision of two black holes with orbital
angular momentum.

In a future work we will apply this construction to the calculations of the
mentioned estimates.

\section*{Acknowledgments}
We acknowledge support from  Fundaci\'on YPF, 
 SeCyT-UNC, CONICET and FONCYT BID 802/OC-AR PICT: 00223.


\begin{thebibliography}{10}

\bibitem{Anninos98}
P.~Anninos and S.~Brandt.
\newblock Head-on collision of two unequal mass black holes.
\newblock {\em Phys.Rev.Lett.}, 81:508, 1998.

\bibitem{Anninos93}
P.~Anninos, D.~Hobill, E.~Seidel, L.~Smarr, and W.-M. Suen.
\newblock Collision of two black holes.
\newblock {\em Phys.Rev.Lett.}, 71:2851, 1993.

\bibitem{Chandra79}
S.~Chandrasekhar.
\newblock An introduction to the theory of the {K}err metric and its
  perturbation.
\newblock In {\em General Relativity, An Einstein Centenary Survey}. Cambridge
  University Press, Cambridge, 1979.
\newblock Edited by S.W. Hawking and W. Israel.

\bibitem{Chandra92}
S.~Chandrasekhar.
\newblock {\em The Mathematica Theory of Black Holes}.
\newblock Oxford University Press, 1992.

\bibitem{Dain96}
S.~Dain, O.~M. Moreschi, and R.~J. Gleiser.
\newblock Robinson-trautman geometries and the photon rockets.
\newblock {\em Class. Quantum Grav.}, 13(5):1155--1160, 1996.

\bibitem{Frittelli92}
S.~Frittelli and O.~M. Moreschi.
\newblock Study of the {R}obinson-{T}rautman metrics in the asymptotic future.
\newblock {\em Gen. Rel. Grav.}, 24:575, 1992.

\bibitem{Geroch73}
R.~Geroch, A.~Held, and R.~Penrose.
\newblock A space-time calculus based on pairs of null directions.
\newblock {\em J. Math. Phys.}, 14:874--881, 1973.

\bibitem{Kramer80}
D.~Kramer, H.~Stephani, MacCallum, and E.~Herlt.
\newblock {\em Exact solutions of Einstein's field equations}.
\newblock Cambridge University Press, Cambridge, 1980.

\bibitem{Moreschi86}
O.~M. Moreschi.
\newblock On angular momentum at future null infinity.
\newblock {\em Class. Quantum. Grav.}, 3:503--525, 1986.

\bibitem{Moreschi87}
O.~M. Moreschi.
\newblock General future asymptotically flat spacetimes.
\newblock {\em Class. Quantum Grav.}, 4:1063--1084, 1987.

\bibitem{Moreschi88}
O.~M. Moreschi.
\newblock Supercenter of mass system at future null infinity.
\newblock {\em Class. Quantum Grav.}, 5:423--435, 1988.

\bibitem{Moreschi99}
O.~M. Moreschi.
\newblock Total energy radiated in the head-on black hole collision with
  arbitrary mass ratio.
\newblock {\em Phys. Rev. D}, 58:084018, 1999.

\bibitem{Moreschi96}
O.~M. Moreschi and S.~Dain.
\newblock Estimates of the total gravitational radiation in the head-on black
  hole collision.
\newblock {\em Phys. Rev. D}, 53(4):R1745--R1749, 1996.
\newblock Rapid Communication.

\bibitem{Moreschi98}
O.~M. Moreschi and S.~Dain.
\newblock Rest frame system for asymptotically flat space-times.
\newblock {\em J. Math. Phys.}, 39(12):6631--6650, 1998.

\bibitem{Moreschi01}
O.~M. Moreschi, A.~Perez, and L.~Lehner.
\newblock Total energy radiated for non head-on binary black hole collisions.
\newblock in preparation, 2001.

\bibitem{Perez01}
A.~Perez and O.~M. Moreschi.
\newblock Characterizing exact solutions from asymptotic physical concepts.
\newblock submited for publication, http://xxx.lanl.gov/ps/gr-qc/0012100, 2000.

\bibitem{Robinson62}
I.~Robinson and R.~Trautman.
\newblock Some spherical gravitational waves in general relativity.
\newblock {\em Proc. R. Soc. A}, 265:463--473, 1962.

\end{thebibliography}

\end{document}